\documentclass{ws-procs9x6}
\makeatletter
\def\alt{\mathrel{\mathpalette\gl@align<}}
\def\agt{\mathrel{\mathpalette\gl@align>}}
\def\gl@align#1#2{\lower.6ex\vbox{\baselineskip\z@skip\lineskip\z@
    \ialign{$\m@th#1\hfil##\hfil$\crcr#2\crcr\sim\crcr}}} \makeatother

\def \gtsim    {\relax\ifmmode{\mathrel{\mathpalette\oversim >}}
  \else{$\mathrel{\mathpalette\oversim >}$}\fi}
\def \ltsim    {\relax\ifmmode{\mathrel{\mathpalette\oversim <}}
  \else{$\mathrel{\mathpalette\oversim <}$}\fi}
\def\oversim#1#2{\lower4pt\vbox{\baselineskip0pt \lineskip1.5pt
    \ialign{$\mathsurround=0pt#1\hfil##\hfil$\crcr#2\crcr\sim\crcr}}}

\def\Journal#1#2#3#4{{#1} {\bf #2}, #3 (#4)}

\begin{document}

\title{Cosmology and Dark Matter at the LHC}

\author{Richard Arnowitt,$^1$ Adam Aurisano,$^1$ Bhaskar Dutta,$^1$ Teruki Kamon,$^1$
\\ Nikolay Kolev,$^2$ Paul Simeon,$^1$ David Toback$^1$ and Peter Wagner$^1$}

\address{$^1$ Department of Physics, Texas A\&M University, College Station, TX 77843-4242, USA\\
 $^2$Department of Physics, University of Regina, Regina, SK S4S
0A2, Canada }

\begin{abstract}
We examine the question of whether neutralinos produced at the LHC can be shown to be the
particles making up the astronomically observed dark matter. If the WIMP alllowed region lies
in the SUGRA coannihilation region, then a strong signal for this would be the unexpected
near degeneracy of the stau and neutralino i.e., a mass difference  $\Delta M\simeq (5-15)$ GeV. For the
mSUGRA model we show such a small mass difference can be measured at the LHC using the signal
$3\tau$+jet+$E_T^{\rm miss}$. Two observables, opposite sign minus like sign pairs and the
peak of the $\tau\tau$ mass distribution allows the simultaneous determination of $\Delta M$
to 15\% and the gluino mass $M_{\tilde g}$ to be 6\% at the benchmark point of $M_{\tilde
g}$=850 GeV, $A_0$=0, $\mu>$0 with 30 fb$^{-1}$. With 10 fb$^{-1}$, $\Delta M$ can be
determined to 22\% and one can probe the parameter space up to $m_{1/2}$=700 GeV with 100
fb$^{-1}$.
\end{abstract}


\bodymatter

\section{Introduction}Supersymmetry (SUSY) offers the possibility of solving a number of
theoretical problems of the Standard Model (SM). Thus the
cancelations implied by the bose-fermi symmetry resolves the gauge
hierarchy problem, allowing one to consider models at energies all
the way up to the GUT or Planck scale. Further, using the SUSY SM
particle spectrum with one pair of Higgs doublets. (pairs of Higgs doublets 
are needed on
theoretical grounds to cancel anomalies and on phenomenological grounds
to give rise to both u and d quark masses) the renormalization group
equations (RGE) show that grand unification of the SM gauge coupling
constants occurs at $M_G\simeq 10^{16}$ GeV opening up the
possibility of SUSY GUT models. However, none of this can actually
occur unless a natural way of spontaneously breaking SUSY occurs, and
this very difficult to do with global supersymmetry. The problem was
resolved by promoting supersymmetry to a gauge symmetry,
supergravity (SUGRA)\cite{a1}, where spontaneous breaking of
supersymmetry can easily occur. One can then build SUGRA GUT models
\cite{a2,a3} with gravity playing a key role in the construction. A
positive consequence of this promotion was that the RGE then show
that the breaking of supersymmetry at the GUT scale naturally leads
to the required $SU(2)\times U(1)$ breaking at the electroweak
scale, thus incorporating all the successes of the SM, without any
prior assumptions of negative (mass)$^2$ terms.

In spite of the theoretical successes of SUGRA GUTs, there has been
no experimental evidence for its validity except for the
verification of grand unification (which has in fact withstood the
test of time for over a decade). However, one expects SUSY particles
to be copiously produced at the LHC. Further, models with R parity
invariance predict the lightest neutralino $\tilde\chi^0_1$ to be a
candidate for the astronomically observed dark matter (DM) and models
exist~\cite{a4} consistent with the amount of dark matter observed
by WMAP~\cite{a5}, and being searched for in the Milky Way by dark
matter detectors. Thus it is possible to build models that both
cover the entire energy range from the electroweak scale to the GUT
scale and go back in time to $\sim 10^{-7}$ seconds after the Big
Bang when the current relic dark matter was created.

The question then arises can we verify if the dark matter particles
in the galaxy is the neutralino expected to be produced at the LHC?
In principle this is doable. Thus assuming the DM detectors
eventually detect the dark matter particle, they will measure the
mass and cross sections, and these can be compared with those
measured at the LHC. However, this may take a long time to achieve.
More immediately, can we look for a signal at the LHC that is
reasonably direct consequence of the assumption that the neutralino
is the astronomical DM particle and in this way experimentally unify
particle phenomena with early universe cosmology? To investigate
this question it is necessary to choose a specific SUSY model, and
for simplicity we consider here the minimal mSUGRA (though a similar
analysis could be done for a wide range of other SUGRA models).

\begin{figure}
\centerline{ \epsfxsize=10.6cm\epsfysize=7.0cm
\epsfbox{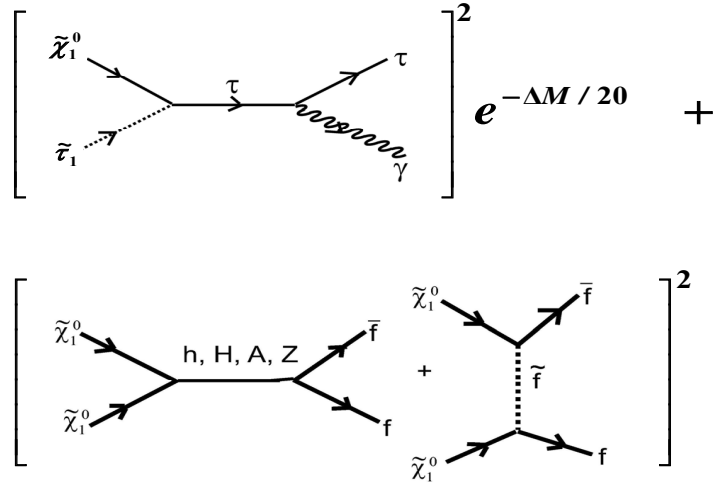}}
\caption{The feynman diagrams for annihilation of neutralino dark matter in the
early universe} \label{fig0}
\end{figure}

\section{The mSUGRA model}The mSUGRA model depends on four soft
breaking parameters  and one sign. These are; $m_{1/2}$ (the
universal gaugino soft breaking mass at $M_G$); $m_0$ (the universal
scalar soft breaking mass at $M_G$); $A_0$ (the universal cubic soft
breaking mass at $M_G$); $\tan\beta=<H_2>/<H_1>$ at the electroweak
scale (where $<H_2>$ gives rise to u quark masses and $<H_1>$ to d
quark masses); and the sign of $\mu$ parameter (where $\mu$ appears
in the quadratic part of the superpotential $W^{(2)}=\mu H_1 H_2$).

Current experimental data significantly constrains these parameters.
the main accelerator constraints are: The Higgs mass $m_H>114$
GeV\cite{a6}; the lightest chargino mass $M_{\tilde\chi^{\pm}_1}>104$ GeV;
 the $b\rightarrow s\gamma$ branching ratio 2.2
$\times 10^{-4}<Br(b\rightarrow s\gamma)<4.5\times 10^{-4}$~\cite{a7}; and the muon
g-2 anomaly~\cite{a8} which now deviates from the SM prediction by 3.4
$\sigma$. The astronomical constraint is the WMAP
determination of the amount of dark matter and we use here a 2
$\sigma$ range~\cite{a5}:
\begin{equation}
0.094<\Omega_{\tilde\chi^0_1}h^2<0.129.\end{equation}
The
WMAP constraint limits the parameter space to three main regions
arising from the diagrams of Fig.1. (1) The stau-neutralino
($\tilde\tau_1-\tilde\chi^0_1$) coannihilation region. Here $m_0$ is
small and $m_{1/2}\leq 1.5$ TeV. (2)The focus region where the
neutralino has a large Higgsino component. Here $m_{1/2}$ is small
and $m_0\geq 1$ TeV. (3) The funnel region where annihilation
proceeds through heavy Higgs bosons which have become relatively
light. Here both $m_0$ and $m_{1/2}$ are large. \vspace*{0.5cm}
\begin{figure}
\centerline{\psfig{file=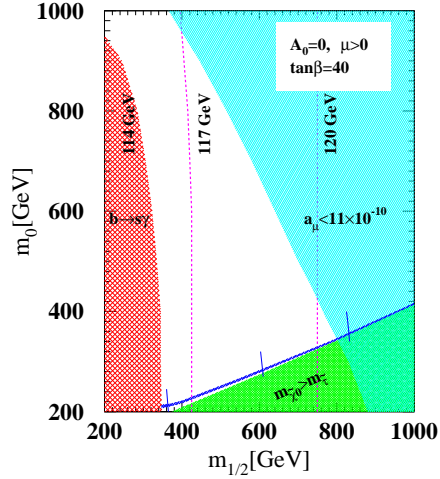,width=2.4in}}
\caption{Allowed parameter space for $\tan\beta$ = 40 with  $A_0
= 0$ and $\mu>0$.} \label{WMAP_allowed_region}
\end{figure}

\begin{figure}
{\epsfxsize=11cm\epsfysize=4.5cm
\epsfbox{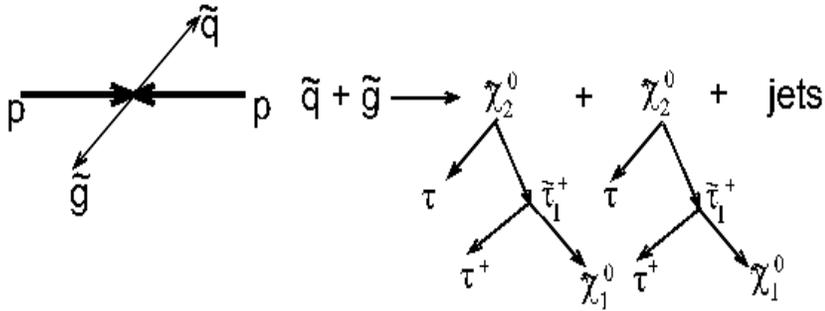}}
\caption{SUSY production and decay channels} \label{fig1}
\end{figure}

(In addition there is a small bulk region) Note that a key element
in the coannihilation region is the Boltzman factor from the
annihilation in the early universe at $kT\sim$ 20 GeV: exp${[-\Delta
M/20]}$ where $\Delta M=M_{\tilde\tau_1}-M_{\tilde\chi^0_1}$. Thus
significant coannihilation occurs provided $\Delta M\leq$ 20 GeV.

The accelerator constraints further restrict the parameter space and
if the muon g-2 anomaly maintains, $\mu>0$ is prefereed and there
remains mainly the coannihilation region. This is illustrated in
Fig.2  which shows  the allowed narrow coannihilation band (for the
case $\tan\beta=40$, $A_0=0$, $\mu>0$) where $\Delta M=(5-15)$ GeV
and $m_{1/2}\leq 800$ GeV. (There is a small focus region for small
$m_{1/2}$ and $m_0>1$ TeV since the $b\rightarrow s \gamma$
constraint ceases to opperate at $m_0>1$ TeV.)

The coannihilation band is narrow ($\Delta M=5-15$ GeV) due to the
Boltzman factor in Fig.1, the range in $\Delta M$ corresponding to
the allowed WMAP range for $\Omega_{\tilde\chi^0_1}h^2$. The dashed
verticle lines are possible Higgs masses.

One may ask two questions; (1) Can such a small stau-neutralino mass
difference (5-15 GeV) arise in mSUGRA, i.e. one would naturally
expect these SUSY particles to be hundreds of GeV apart and (2) Can
such a small mass difference be measured at the LHC? If the answers
to both  these questions are affirmative, the observation of such a
small mass difference would be a strong indication that the
neutralino is the astronomical DM particle since it is the
cosmological constraint on the amount of DM that forces the near
mass degeneracy with the stau, and it is the accelerator constraints
that suggests that the coannihilation region is the allowed region.

\section{Can $\Delta M$ be Small in SUGRA models?}
At the GUT scale $m_{1/2}$ governs  the gaugino masses, while $m_0$
the slepton masses. Thus, at $M_G$ one would not expect any
degeneracies between the two classes of particles. However, at the
electroweak scale the RGE can modify this result. To see
analytically this possibility , consider the lightest selectron
$\tilde e^c$ which at the electroweak scale has mass
\begin{equation}
m^2_{\tilde e^c}=m_0^2+0.15 m_{1/2}^2+(37 \rm GeV)^2\end{equation}
while the $\tilde\chi^0_1$ has mass

\begin{equation}
m^2_{\tilde \chi^0_1}=0.16 m_{1/2}^2 \end{equation}
The numerical accident that the coefficients of $m_{1/2}^2$ is
nearly the same for both cases allows a near degeneracy. Thus for
$m_0=0$, the $\tilde e^c$ and $\tilde\chi^0_1$ become degenerate at
$m_{1/2}$=(370-400) GeV. For larger $m_{1/2}$, the near degeneracy is
maintained by increasing $m_0$, so that one can get the narrow
corridor in the $m_0$-$m_{1/2}$ plane seen in Fig.2. Actually the
case of the stau $\tilde\tau_1$ is more complicated since the large
t-quark mass causes left-right mixing in the stau mass matrix and
results in the $\tilde\tau_1$ being the lightest slepton (not the
selectron). However, a result similar to Eqs. (1,2) occurs, with a
$\tilde\tau_1-\tilde\chi^0_1$ coannihilation corridor resulting.

We note that the results of Eqs.(1,2) depend only on the U(1) gauge
group and so coannihilation can occur even if there were
non-universal scalar mass soft-breaking or non-universal gaugino
mass soft breaking at $M_G$. Thus, coannihilation can occur in a
wide class of SUGRA models, and is not just a feature of mSUGRA.

\section{Coannihilation signal at the LHC}

At the LHC, the major SUSY production processes are gluinos ($\tilde
g$) and squarks ($\tilde q$) e.g., $p+p\rightarrow \tilde g+\tilde
q$. These then decay into lighter SUSY particles and Fig.3 shows a
major decay scheme. The final states involve two $\tilde\chi^0_1$
giving rise to missing transverse energy $E^T_{\rm miss}$) and four
$\tau$'s, two from the $\tilde g$ and two from the $\tilde q$ decay chain for the
example of Fig 3. In the
coannihilation region, two of the taus are high energy (``hard" taus)
coming from the $\tilde\chi^0_2\rightarrow\tau\tilde\tau_1$ decay
(since $M_{\tilde\chi^0_2}\simeq 2 M_{\tilde\tau_1}$) while two are
low energy (``soft" taus) coming from the
$\tilde\tau_1\rightarrow\tau+\tilde\chi^0_1$ decay since $\Delta M$ is
small. The signal is thus $E_T^{\rm miss}+$ jets +$\tau$'s, which
should be observable at the LHC detectors.

\begin{figure}
{\epsfxsize=6cm\epsfysize=6.0cm
\epsfbox{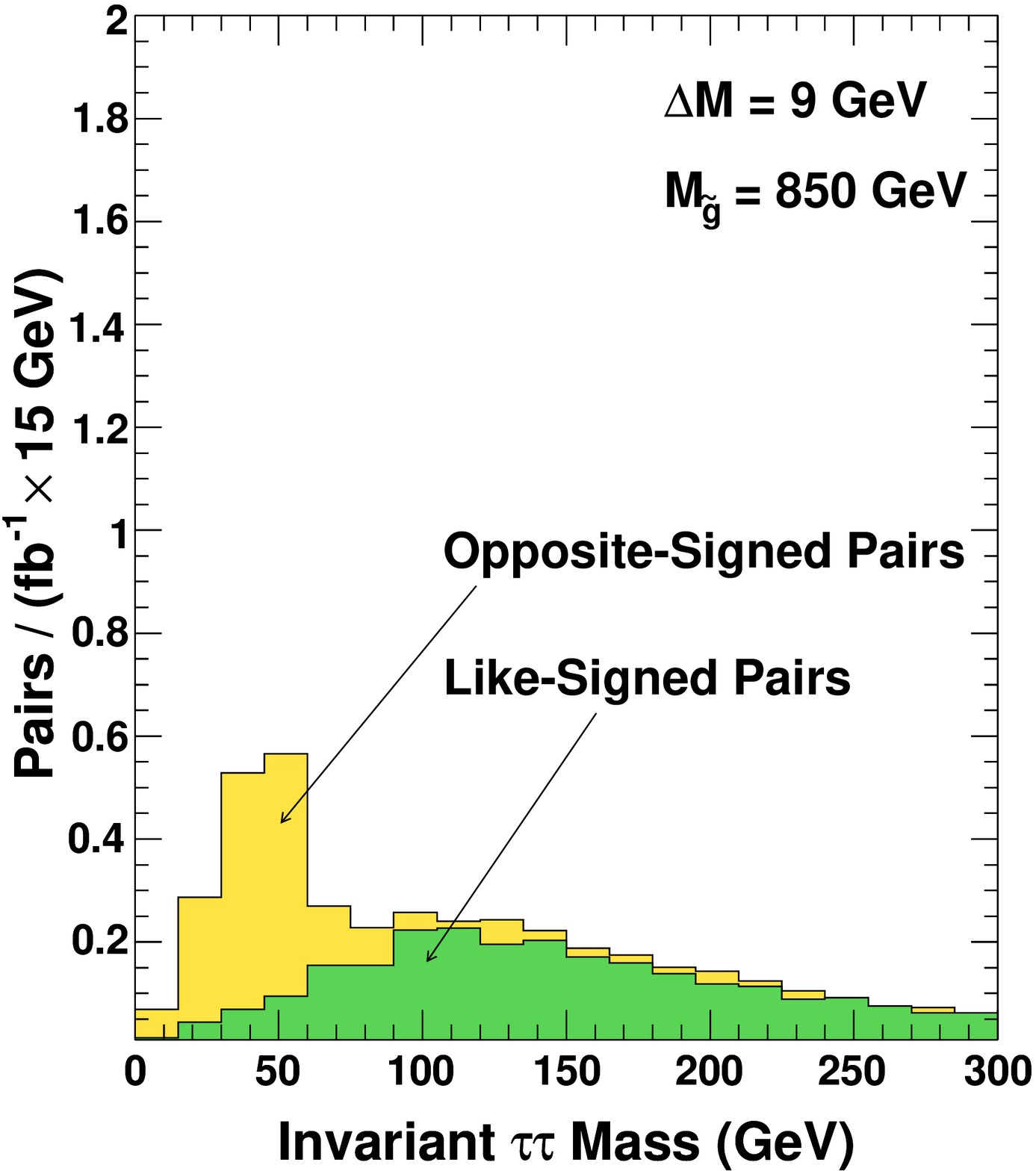}}
{\epsfxsize=6cm\epsfysize=6.0cm
\epsfbox{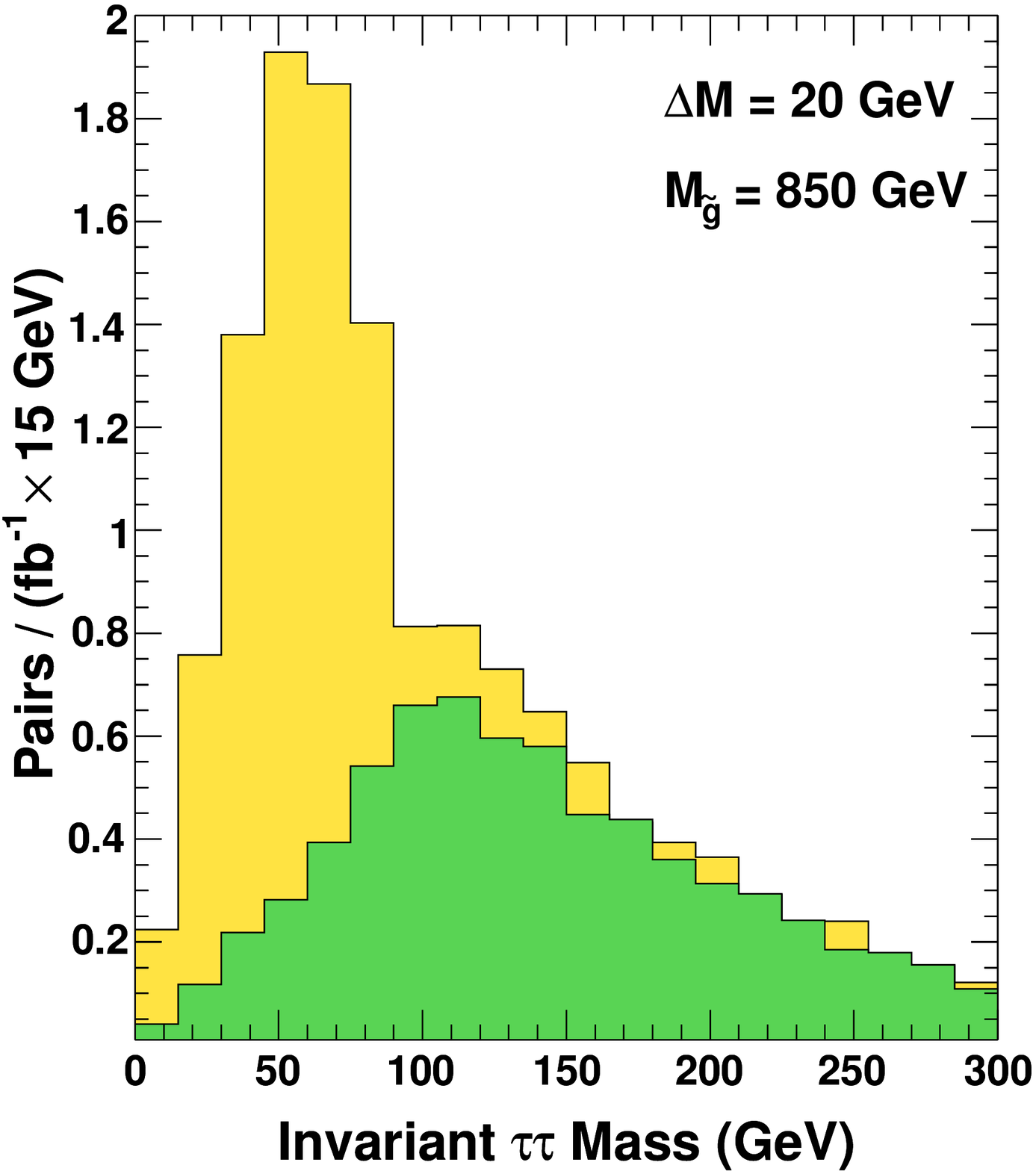}}
\caption{Number of tau pairs as a function of invariant $\tau\tau$
mass. The difference $N_{OS}$-$N_{LS}$ cancels for mass $\geq$ 100
GeV eliminating background events}\label{fig2}
\end{figure}

As seen above we expect two pairs of taus, each pair containing one
soft and one hard tau from each $\tilde\chi^0_2$ decay. Since
$\tilde\chi^0_2$ is neutral, each pair should be of opposite
sign(while SM and SUSY backgrounds, jets faking taus will have equal
number of like sign as opposite sign events). Thus one can suppress
backgrounds statistically by considering the number of oppsite sign
events $N_{OS}$ minus the like sign events $N_{LS}$. The four $\tau$
final state has the smallest background but the acceptance and
efficiency for reconstructing all four taus is low. Thus to
implement the above ideas we consider here the three $\tau$ final
state of which two are hard  and one is soft. (The two $\tau$ final
state with higher acceptance but larger backgrounds was discussed in
\cite{a91}, and an analysis of the coannhilation signal at the ILC
was given in \cite{a10}.

We label three taus by their transverse energies with
$E^T_1>E^T_2>E^T_3$ and form the pairs 13 and 23. For signal events
one of the two pairs should be coming from a $\tilde\chi^0_2$ decay
and have opposite sign(OS) while the other is not correlated. There
are two measurables that can be formed. The number $N$ and the mass
of the pair $M$. To simulate the data we use ISAJET 7.64 \cite{a11}
and PGS detector simulator \cite{a12}. Events are chosen with
$E_T^{\rm miss}$ and 1 jet and three taus with visible momenta
$p_T^{\rm vis}>40$ GeV, $p_T^{\rm vis}>40$ GeV, $p_T^{\rm vis}>20$
GeV. We assume here that it is possible to reconstruct taus with
$p_T^{\rm vis}$ as low as 20 GeV. Standard Model background is
reduced by requiring $E^{\rm jet1}_T>100$ GeV, $E^{\rm miss}_T>100$
GeV with tevatron results, $E^{\rm jet1}_T+E^{\rm miss}_T>400$ GeV. We also assume rate of
a jet faking a $\tau$ ($f_{j\rightarrow\tau}$) to be
$f_{j\rightarrow\tau}=1\%$ (with a 20\% error in
$f_{j\rightarrow\tau}$) consistent with Tevatron results.

Fig 4. shows the number of events as a function of the $\tau\tau$
mass for gluino mass $M_{\tilde g}=850$ GeV and $\Delta M=9$ GeV and
20 GeV. One sees that the difference $N_{OS}-N_{LS}$ cancels out as
expected for $\tau\tau$ mass $\geq$100 GeV (consistent with the fact
that the signal events are expected to lie below 100 GeV).

Fig. 5 shows the behavior of $N_{OS-LS}$ as a function of $\Delta M$
and $M_{\tilde g}$. The central black line is for the assumed 1\%.
rate for jets faking a $\tau$, the shaded region around it is for a
20\% uncertainty in $f_{j\rightarrow\tau}$. One sees that provided
this uncertainty is not large, it produces only a small effect.

\begin{figure}
{\epsfxsize=6cm\epsfysize=6.0cm
\epsfbox{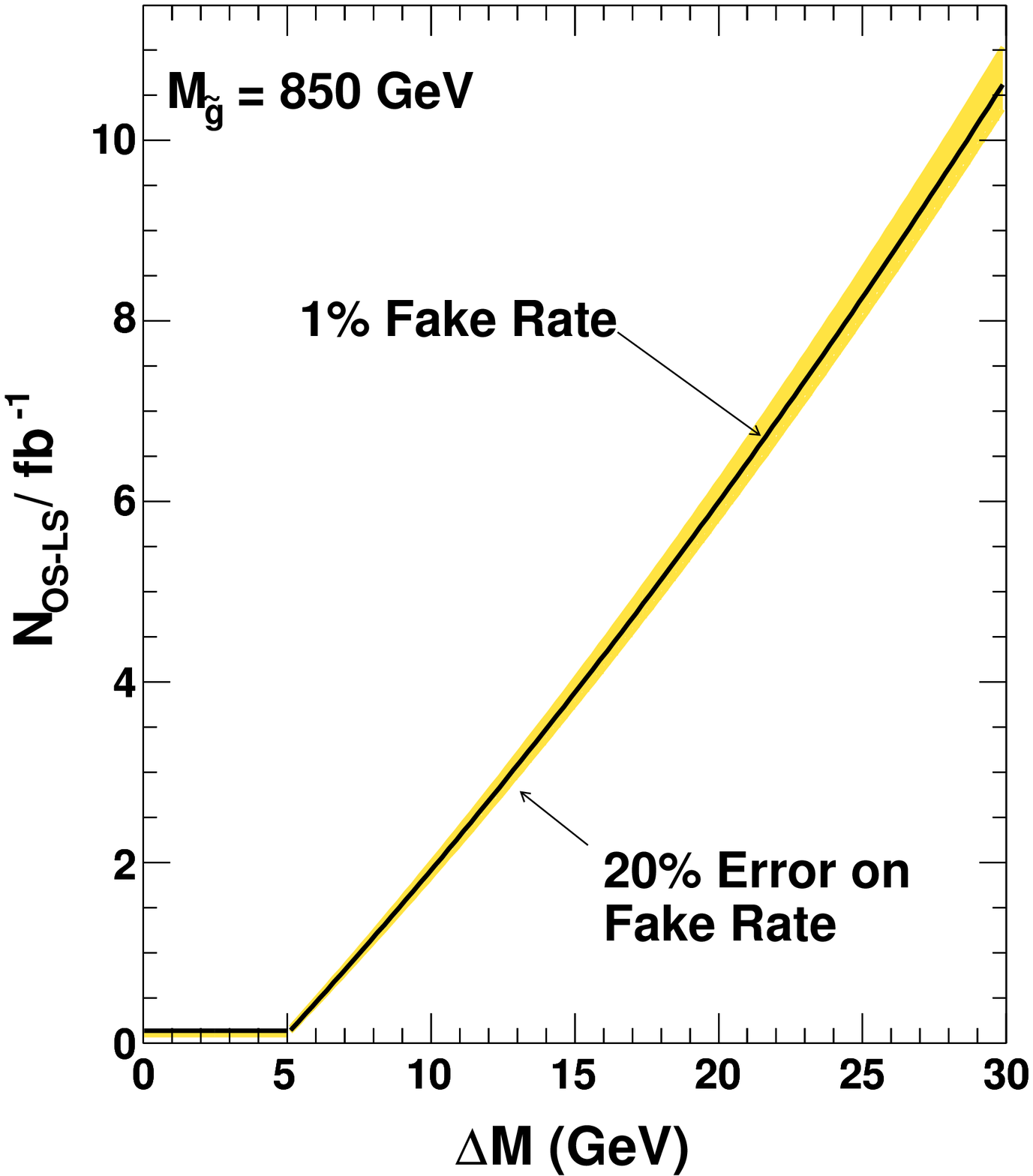}}
{\epsfxsize=6cm\epsfysize=6.0cm
\epsfbox{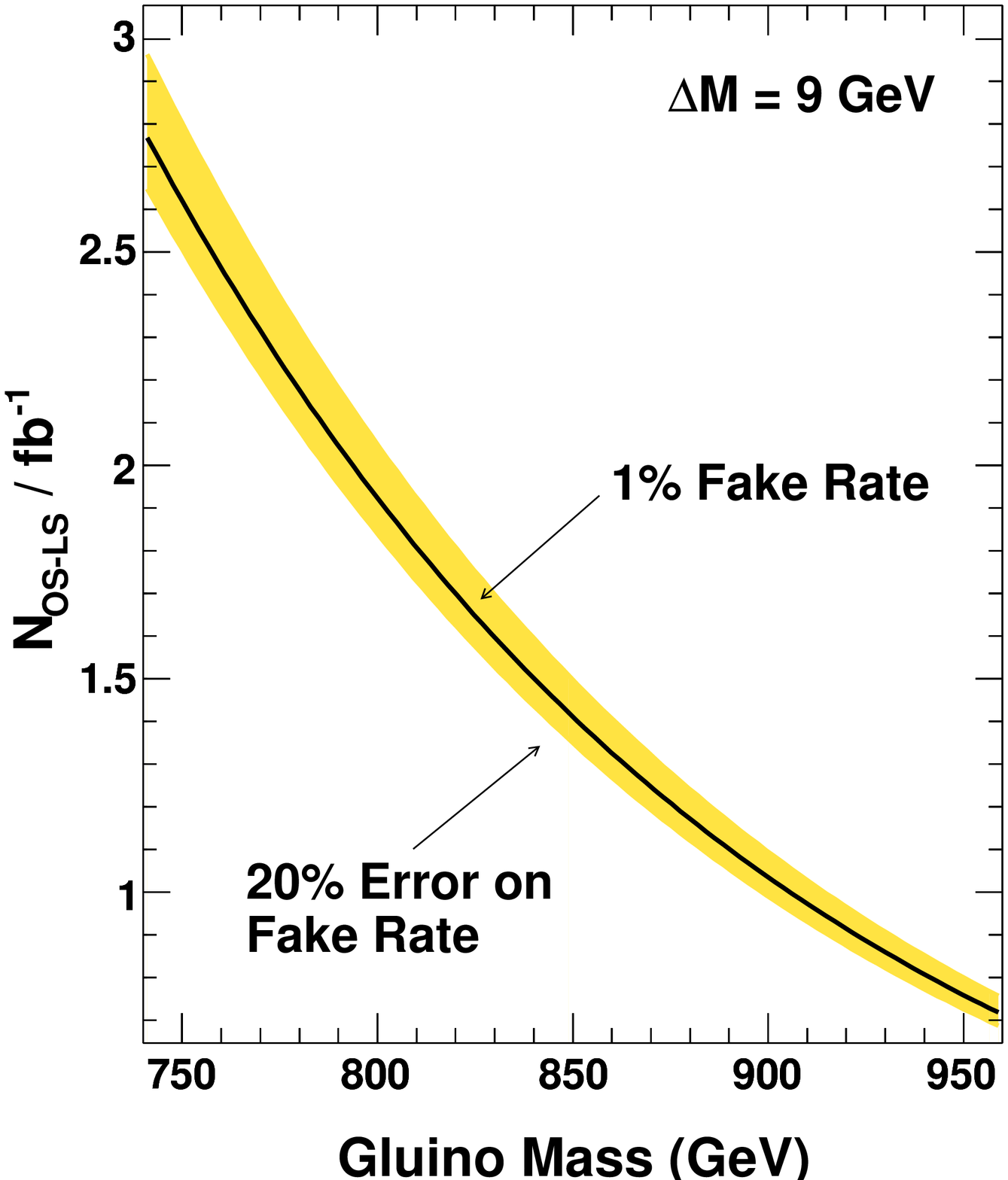}}
\caption{$N_{OS-LS}$ as function of $\Delta M$ (left graph) and as a
function of $M_{\tilde g}$ (right graph). The central black line
assumes a 1\% fake rate, the shaded area representing the 20\% error
in the fake rate}\label{fig3}
\end{figure}

Figs 4 and 5 show two important features. First, $N_{OS-LS}$
increases with $\Delta M$(since the $\tau$ acceptance increases) and
$N_{OS-LS}$ decreases with $M_{\tilde g}$(since the production cross
section of gluinos and squarks decrease with $M_{\tilde g}$).
Second, from Fig.4 one sees that $N_{OS-LS}$ forms a peaked
distribution \cite{a91,a13}. The ditau peak position $M_{\tau\tau}^{\rm
peak}$ increases with both $\Delta M$ and $M_{\tilde g}$. This
allows us to use the two measurables $N_{OS-LS}$ and
$M_{\tau\tau}^{\rm peak}$ to determine both $\Delta M$ and
$M_{\tilde g}$. Fig.6 shows this determination for the benchmark
case of $\Delta M$=9 GeV, $M_{\tilde g}$=850 GeV, $A_0$=0 and
$\tan\beta$=40. Plotted there are constant values of $N_{OS-LS}$ and
constant values of $M_{\tau\tau}^{\rm peak}$ in the $\Delta
M-M_{\tilde g}$ plane which exhibit the above dependance of these
quantities on $\Delta M$ and $M_{\tilde g}$. With luminosity of 30
fb$^{-1}$ one determines $\Delta M$ and $M_{\tilde g}$ with the
following accuracy:
\begin{equation}
\delta \Delta M/\Delta M\simeq 15\%;\,\, \delta M_{\tilde
g}/M_{\tilde g}=6\%
\end{equation}
Fig. 7 shows how the accuracy of the measurement changes with
luminosity. One sees that even with 10 fb$^{-1}$ (which should be
available at the LHC after about two years running)one could
determine $\Delta M$ to within 22\%, which should be sufficient to
know whether one is in the SUGRA coannihilation region.
\begin{figure}
\centerline{\psfig{file=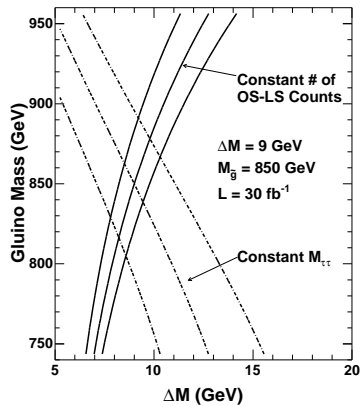,width=2in}}
\caption{Simultaneous determination of $\Delta M$ and $M_{\tilde
g}$. The three lines plot constant $N_{OS-LS}$ and $M_{\tau\tau}^{\rm
peak}$ (central value and 1$\sigma$ deviation) in the $M_{\tilde
g}$-$\Delta M$ plane for the benchmark point of $\Delta M$=9 GeV and
$M_{\tilde g}$=850 GeV assuming 30 fb$^{-1}$ luminosity}
\label{fig4}
\end{figure}

\begin{figure}
\centerline{\psfig{file=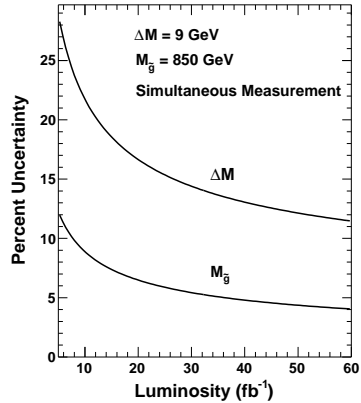,width=2in}}
\caption{Uncertainty in the determination of $\Delta M$ and
$M_{\tilde g}$ as a function of luminosity.} \label{fig5}
\end{figure}

\section{Conclusions}
We have examined here the question of how one might show that the
$\tilde\chi^0_1$ particle produced at the LHC is the astronomically
observed dark matter. If $\Delta M$, the stau-neutralino mass
difference lies in the coannihilation region of the SUGRA
$m_0$-$m_{1/2}$ plane where $\Delta M=(5-15)$ GeV, this would be
strong indication that the neutralino is the dark matter particle as
otherwise the mass difference would not naturally be so small. We
saw how it was possible to measure such a small mass difference at
the LHC for the mSUGRA model using a signal of $E_T^{\rm miss}$+ 1
jet+$3\tau$, and simultaneously determine the gluino mass $M_{\tilde
g}$, provided it is possible at the LHC to reconstruct taus with
$p_T^{\rm vis}$ as low as 20 GeV. With 30 fb$^{-1}$ one could then
determine $\Delta M$ with 15\% accuracy and $M_{\tilde g}$ with 6\%,
at our benchmark point of $\Delta M$=9 GeV, $M_{\tilde g}$=850 GeV,
$\tan\beta=40$, and $A_0$=0. Even with only 10fb$^{-1}$ one would
determine $\Delta M$ to within 25\% accuracy, sufficient to learn
whether the signal is in the coannihilation region.

While the analysis done here was within the framework of mSUGRA,
similar analyses can be done for other SUGRA models provided the
production of neutralinos is not suppressed. However, the
determination of $M_{\tilde g}$ does depend on the mSUGRA
universality of the gaugino masses at $M_G$ to relate
$M_{\tilde\chi^0_2}$ to $M_{\tilde g}$. Thus a model independent
method of determining $M_{\tilde g}$ would allow one to to test the
question of gaugino universality. However, it may not be easy to
directly measure $M_{\tilde g}$ at the LHC for high $\tan\beta$ in
the coannihilation region due to the large number of low energy
taus, and the ILC would require a very high energy option to see the
gluino.

As mentioned above, one can also measure $\Delta M$ using the signal
$E_T^{\rm miss}$+ 2 jets+2$\tau$ \cite{a91}. This signal has higher
acceptance but larger backgrounds. There, with 10 fb$^{-1}$ one
finds that one can measure $\Delta M$ with 18\% error  at the
benchmark point assuming a separate measurement of $M_{\tilde g}$
with 5\% error has been made. While we have fixed our benchmark
point at $M_{\tilde g}=850$ GeV(i.e. $m_{1/2}=$360 GeV), higher
gluino mass would require more luminosity to see the signal. One finds that with 100 fb$^{-1}$ one can
probe $m_{1/2}$ at the LHC up to $\sim 700$ GeV (i.e., $M_{\tilde
g}$ up to $\simeq 1.6$ TeV).

Finally, it is interesting to compare with possible measurements of
$\Delta M$ at the ILC. If we implement a very forward calorimeter to
reduce the two $\gamma$ background, $\Delta M$ can be determined
with 10\% error at the benchmark point \cite{a10}. Thus in the
coannihilation region, the determination of $\Delta M$ at the LHC is
not significantly worse than at the ILC.

\section{Acknowledgement} This work was supported in part by DOE
grant DE-FG02-95ER40917, NSG grant DMS 0216275 and the Texas A\&M
Graduate Merit fellowship program.


\begin{thebibliography}{9}

\bibitem{a1}D.Z. Freedman, P. Van Niewenhuisen, and S. Ferrara,
\Journal{Phys. Rev.}{D13}{1980}{1729}; S. Deser and B. Zumino,
\Journal{Phys. Lett.}{B65}{369}{1976};\Journal{Phys. Lett.}{B65}
{1976}{369}.

\bibitem{a2}A.H. Chamseddine, R. Arnowitt, and P. Nath,
\Journal{Phys. Rev.Lett}{49}{1982}{970}.

\bibitem{a3}
R. Barbieri, S. Ferrara, and C.A. Savoy,  \Journal{Phys.
Lett.}{B119}{343}{1982}; L. Hall, J. Lykken, and S.
Weinberg,\Journal{Phys. Rev.}{D27}{1983}{2359}; P. Nath, R.
Arnowitt, and A.H. Chamseddine, \Journal{Nucl.
Phys.}{B227}{121}{1983};
 For a review, see P. Nilles, \Journal{Phys. Rept.}{100}{1984}{1}.

\bibitem{a4}
J. Ellis, K. Olive, Y. Santoso, and V. Spanos, \Journal{Phys.
Lett.}{B176}{565}{2003}; R. Arnowitt, B. Dutta, and B. Hu,
hep-ph/0310103; H. Baer, C. Balazs, A. Belyaev, T. Krupovnickas, and
X. Tata, \Journal{JHEP}{0306}{2003}{054}; B. Lahanas and D.V.
Nanopoulos, \Journal{Phys. Lett.}{568}{2003}{55}; U. Chattopadhyay,
A. Corsetti, and P. Nath, \Journal{Phys. Rev.}{D68}(2003){035005};
E. Baltz and P. Gondolo, JHEP {\bf 0410} (2004) 052; B.~C.~Allanach
and C.~G.~Lester, \Journal{Phys. Rev.}{D73}{2006}{015013}; A.
Djouadi, M. Drees, and J-L. Kneur, \Journal{JHEP}{0603}{2006}{033}.

\bibitem{a5}WMAP Collaboration, D.N. Spergel {\it et al.}, \Journal{Astrophys. J. Suppl.}{148}{2003}{175}.




\bibitem{a6}
ALEPH, DELPHI, L3, OPAL Collaborations, G. Abbiendi {\it et al.} (The LEP Working Group for Higgs Boson Searches), Phys. Lett. B {\bf 565} (2003) 61;
Particle Data Group, S.~Eidelman {\it et al.}, \Journal{Phys. Lett.}{592}{2004}{1}.

\bibitem{a7}M. Alam {\it et al.},  \Journal{Phys. Rev. Lett}{74}{1995}{2885}.

\bibitem{a8}
Muon $g-2$ Collaboration, G. Bennett {\it et al.}, \Journal{Phys.
Rev. Lett}{92}{2004}{161802}; S. Eidelman, Talk at ICHEP 2006,
Moscow, Russia.


\bibitem{a91}R. Arnowitt, B. Dutta, T. Kamon, N. Kolev, and D. Toback, \Journal{Phys.
Lett.}{B639}{2006}{46}.

\bibitem{a10} V. Khotilovich, R. Arnowitt, B. Dutta, and T.
Kamon,\Journal{Phys. Lett.}{B618}{2005}{182}.

\bibitem{a11}F. Paige, S. Protopescu, H. Baer, and X. Tata, hep-ph/0312045.  We
use ISAJET version 7.64.

\bibitem{a12}PGS is a parameterized detector simulator.  We used the
CDF detector information to obtain approximate jet finding.
We used version 3.2 (see http://www.physics.ucdavis.edu/\verb+~+conway/research/software/pgs/pgs4-general.htm).

\bibitem{a13}I. Hinchliffe and F.E. Paige, \Journal{Phys. Rev.}{D61}{2000}{095011}.

\end{thebibliography}

\end{document}